\documentclass[amsmath,amssymb,aps,preprintnumbers,twocolumn]{revtex4-1}
\pdfoutput=1
\usepackage{epsf}

\usepackage{epsfig,graphics}
\usepackage {graphicx}
\usepackage {epsfig}
\usepackage {subfigure}
\usepackage {tabularx} 
\usepackage{rotate}	
\usepackage{slashed}

\usepackage{amsmath}
\usepackage{amsfonts}
\usepackage{amssymb}
\usepackage{graphicx}

\newcommand{\bmat}{\left(\begin{array}}
\newcommand{\emat}{\end{array}\right)}

\def\yzero{\smash{\hbox{$y\kern-4pt\raise1pt\hbox{${}^\circ$}$}}}

\def\beq{\begin{equation}}
\def\eeq{\end{equation}}
\def\beqa{\begin{eqnarray}}
\def\eeqa{\end{eqnarray}}

\def\-{\hphantom{-}}

\def\s2{\frac{1}{\sqrt2}}

\def\beq{\begin{equation}}
\def\eeq{\end{equation}}
\def\beqa{\begin{eqnarray}}
\def\eeqa{\end{eqnarray}}

\def\IF{\relax{\rm I\kern-.18em F}}
\def\II{\relax{\rm I\kern-.18em I}}
\def\IP{\relax{\rm I\kern-.18em P}}
\def\IC{\relax\hbox{\kern.25em$\inbar\kern-.3em{\rm C}$}}
\def\IR{\relax{\rm I\kern-.18em R}}

\def\Dsl{\,\raise.15ex\hbox{/}\mkern-13.5mu D} 
\def\IZ{Z\kern-.4em  Z}

\catcode`\@=11   
\newdimen\@rotdimen
\newbox\@rotbox  

\def\@vspec#1{\special{ps:#1}}
\def\@rotstart#1{\@vspec{gsave currentpoint currentpoint translate
   #1 neg exch neg exch translate}}
\def\@rotfinish{\@vspec{currentpoint grestore moveto}}
\def\@rotr#1{\@rotdimen=\ht#1\advance\@rotdimen by\dp#1%
   \hbox to\@rotdimen{\hskip\ht#1\vbox to\wd#1{\@rotstart{90 rotate}%
   \box#1\vss}\hss}\@rotfinish}
\def\@rotl#1{\@rotdimen=\ht#1\advance\@rotdimen by\dp#1%
   \hbox to\@rotdimen{\vbox to\wd#1{\vskip\wd#1\@rotstart{270 rotate}%
   \box#1\vss}\hss}\@rotfinish}%
\def\@rotu#1{\@rotdimen=\ht#1\advance\@rotdimen by\dp#1%
   \hbox to\wd#1{\hskip\wd#1\vbox to\@rotdimen{\vskip\@rotdimen
   \@rotstart{-1 dup scale}\box#1\vss}\hss}\@rotfinish}%
\def\@rotf#1{\hbox to\wd#1{\hskip\wd#1\@rotstart{-1 1 scale}%
   \box#1\hss}\@rotfinish}%
\def\rotate{\@ifnextchar[{\@rotate}{\@rotate[l]}}
\def\@rotate[#1]#2{\setbox\@rotbox=\hbox{#2}\@nameuse{@rot#1}\@rotbox}

\catcode`\@=12

\begin{document}

\preprint{IFT-UAM/CSIC-14-026}

\title{BICEP2, the Higgs Mass and the SUSY-breaking Scale}
\author{Luis E. Ib\'a\~nez}
\author{Irene Valenzuela}
\address{Departamento de F\'{\i}sica Te\'orica 
and Instituto de F\'{\i}sica Te\'orica  UAM-CSIC,\\
Universidad Aut\'onoma de Madrid,
Cantoblanco, 28049 Madrid, Spain}
\begin{abstract}
Recent BICEP2 results on CMB polarisation B-modes suggest a high value for the inflation scale
$V_0^{1/4} \simeq 10^{16}$  GeV, giving experimental
 evidence for a physical scale in between the EW scale and the Planck mass. We propose  that this 
 new high scale
could  be interpreted as evidence for a high SUSY breaking scale $M_{ss}\simeq 10^{12}-10^{13}$
 GeV. We show that such a large value for $M_{ss}$ is consistent 
 with a  Higgs mass around 126 GeV. 
 We briefly discuss some possible particle physics implications of this assumption.
\end{abstract}
\maketitle
{\it Introduction.}
The BICEP2 collaboration has recently reported the measurement of cosmological B-mode
polarisation in the CMB \cite{bicep2}.  The observed tensor to scalar ratio $r=0.20^{+0.07}_{-0.05}$ is unexpectedly large.
Since $r$ is related to the energy scale of inflation $V_0$ by
\beq
V_0^{1/4} \ \simeq \ 2\times 10^{16}\left(\frac {r}{0.20}\right)^{1/4}  \ \text{GeV}  \ 
\label{tomaya}
\eeq
these data  give  first experimental  evidence for  the existence of a new physics scale in between the EW and
Planck scales.  This fact, if indeed confirmed,  has important implications  for particle physics. 
The value of $V_0$ 
could suggest this scale could have something to do with a GUT scale $M_X\simeq 10^{16}$ GeV.
On the other hand, if one thinks that SUSY is a fundamental symmetry of the SM which 
is spontaneously broken at some scale, one could think that  the height of the inflation potential
could  be of the same order as the height of the SUSY breaking scalar potential.
In particular the latter is expected to be of order
\beq
V_{ss} \ \simeq \  (m_{3/2}M_p)^2
\eeq
with $m_{3/2}$ the gravitino mass, which also gives us the typical size of SUSY breaking soft terms.
Then the BICEP2 results could be pointing to a SUSY breaking scale 
\beq 
M_{ss} \ \simeq   \  \frac {V_0^{1/2}}{M_p} \ \simeq \ 10^{13} \ \text{GeV} \ .
\eeq
Specifically, the simplest inflation model in agreement with BICEP2 data is given by the simple
chaotic inflationary model \cite{chaotic} with
\beq
V_I \ = \ \frac {m_I^2}{2}\phi^2 \ .
\eeq
The inflation mass $m_I$  in chaotic inflation,
in which $\phi$ reaches $M_p$ at inflation would be of order $10^{13}$ GeV. 
It could also be slightly lower, $m_I\simeq 10^{12}$ GeV, 
if, e.g., one takes into account possible corrections coming  from dim$>4$   polynomials in $\phi$, see e.g.\cite{veronica}.
In the present scheme, the inflaton mass
parameter could be generated by broken SUSY, suggesting $M_{ss}\simeq m_I\simeq 10^{12}-10^{13}$ GeV.
Note that we are not claiming that  low energy SUSY, with soft terms at the  TeV scale is in conflict with
the BICEP2 results. Only that it would require  the height of the inflaton potential to be much higher that
the SUSY-breaking scalar potential. This may lead to problems e.g. in scenarios in which there are
moduli whose vevs are fixed upon SUSY breaking, see e.g.\cite{kl} (see also e.g.\cite{aggj} for other inflationary schemes 
with large SUSY breaking).

In what follows we will  assume that the BICEP2 results are indeed pointing to a SUSY-breaking scale
$M_{ss}\simeq 10^{12}-10^{13}$ GeV and derive some consequences. In particular its consistency 
with the observed Higgs mass value. See \cite{manybicep,bicepaxions} for other recent papers on implications of the
BICEP2 results.

{\it Intermediate SUSY breaking scale and Higgs mass.}
If this new scale is present, 
the EW  hierarchy problem becomes even more pressing, since loops involving the
heavy states associated to the new scale  will presumably give quadratic large contributions to the Higgs mass
which cannot be ignored. 
On the other hand, with SUSY broken at such high scales \cite{hn} , it will not be relevant for the solution to the hierarchy 
problem. It seems then that the Higgs mass should be somehow fine-tuned to survive 
at low-energies.
 There are however 
additional reasons to believe that a SUSY extension of the SM could still apply at some energy scale,
though possibly a very large one.  In particular, supersymmetry is a built-in symmetry inside
string theory, which is the leading candidate for an ultraviolet completion of the SM, including
gravity. Also a SUSY version of the standard model guarantees stabilty (absence of tachyons) for 
the abundant scalars appearing in generic string compactifications. On the other hand the existence of a 
string landscape may provide a rationale for understanding the origin of fine-tuning.

Irrespective of any 
string theory arguments,  having SUSY at some (possibly large)  scale  may solve the stability problem of
the Higgs scalar potential. Indeed, if one extrapolates the value of the  Higgs SM self-coupling $\lambda$ 
up in energies according to the RGE, the top quark loops make  it to vanish and then become negative  at scales of order 
$10^{11}-10^{12}$ GeV, signalling an instability (or metastability)  of the Higgs scalar potential at very high energies
\cite{previos,elias}.
If SUSY is restored around those energies, the Higgs potential is automatically stabilised, since a SUSY
potential is always positive definite. 
In addition to stabilising the Higgs vacuum, high scale SUSY breaking may provide an understanding of the
observed value of the Higgs mass \cite{hebecker1,imrv,iv,hebecker2} (see also \cite{otrosinter}).
 Specifically, in   \cite{iv} it  was shown that,
a SUSY breaking scale  above $10^{10}$ GeV generically gives rise to values $m_H=126\pm 3$ GeV for
the Higgs.  Let us see how this comes about. 
Let us assume for simplicity that above a large SUSY-breaking scale scale $M_{ss}$  one recovers the
MSSM structure. The Higgs sector then has a general mass matrix
\small
 \beq
\left(
\begin{array}{cc}
 {{H_u}} \!\!\! & ,\  {{ H}_d^*}\\
\end{array}
\right)
\left(
\begin{array}{cc}
 { m_{H_u}^2(Q)}+\mu^2(Q) &   m_{3}^2(Q)\\   
  { m_3^2(Q)} & {m_{H_d}^2(Q)+\mu^2(Q) }\\
\end{array}
\right)
\left(
\begin{array}{c}
  {{ H_u^*}} \\
  {{ H}_d}\\
\end{array}
\right) 
\label{matrizmasas}
\eeq
\normalsize
where $Q$ is the running scale, and $\mu$ is a standard MSSM mu-term. 
 For  a SM  Higgs boson to remain light  below the $M_{ss}$ scale one has to fine-tune
\beq
det\left({M_H^2(M_{ss})}\right)\  =\ 0 \ .
\eeq
This could happen if, at a unification scale $Q=M_X$ the mass matrix has only positive eigenvalues and then 
at the lower running scale $Q=M_{ss}$  the determinant vanishes.  
The fine-tuning  condition is
\beq
 (m_{H_u}^2(M_{ss})+\mu^2(M_{ss}))( m_{H_d}^2(M_{ss})+\mu^2(M_{ss}))=m_3^4(M_{ss}) 
 \eeq
  and then one can check that  the linear combination
$H_{SM} = sin\beta H_u + cos\beta H_d^* $ remains light and becomes the SM Higgs field. 
Here the mixing angle is given by
\beq
 tan\beta (M_{ss})\ =\    \left|  \frac {m^2_{H_d}(M_{ss})+\mu^2(M_{ss})}{m^2_{H_u}(M_{ss})+\mu^2(M_{ss})}\right|^{1/2} \ ,
 \eeq
  while 
the Higgs self-coupling at $M_{ss}$  is given by the $MSSM$ boundary condition
\cite{split,hn}
\beq
\lambda_{SUSY}(M_{ss}) \ =\ \frac {1}{4}
(g_2^2(M_{ss}) \ +\  g_1^2(M_{ss})) \ cos^22\beta(M_{ss}) \ .
\label{lambdasusy}
\eeq
A natural additional condition to impose is that 
 $m_{H_u} = m_{H_d}$ at the unification scale $M_X$. 
 This happens in a variety of models including most GUT's and string theory
 frameworks.
 Note  that one then has $tan\beta =1$ at the unification
 scale,  but it runs to a value $tan\beta>1$ at $M_{ss}$. Still  $cos^22\beta$ remains  small, explaining why 
 the Higgs self-coupling is close to zero at scales $M_{ss}> 10^{10}$ GeV.
 One can compute the value of $tan\beta$ at $M_{ss}$ by running it down to the $M_{ss}$ scale.
The computation turns out to be quite independent on the choice of soft terms for the running,
as long as they are all of the same order of magnitude. There is a mild dependence on the $\mu$ parameter 
that we will show explicitly below. 
One can then obtain the value of the self-coupling in eq.(\ref{lambdasusy}) by inserting the value 
so obtained for $\beta$. 
 The EW gauge couplings  $g_{1,2}^2(M_{ss})$ are obtained running up  their experimental value from the EW scale. 
Once we know the value of $\lambda_{SUSY}(M_{ss})$, one can then finally  run it down to the EW scale 
and compute the Higgs mass from 
$m_H^2(Q_{EW})\ =\ 2 v^2(\lambda(Q_{EW}))$, see \cite{iv} for the relevant RGE, thresholds and other details.

\begin{figure}[t]
\begin{center}
\includegraphics[width=9.2cm]{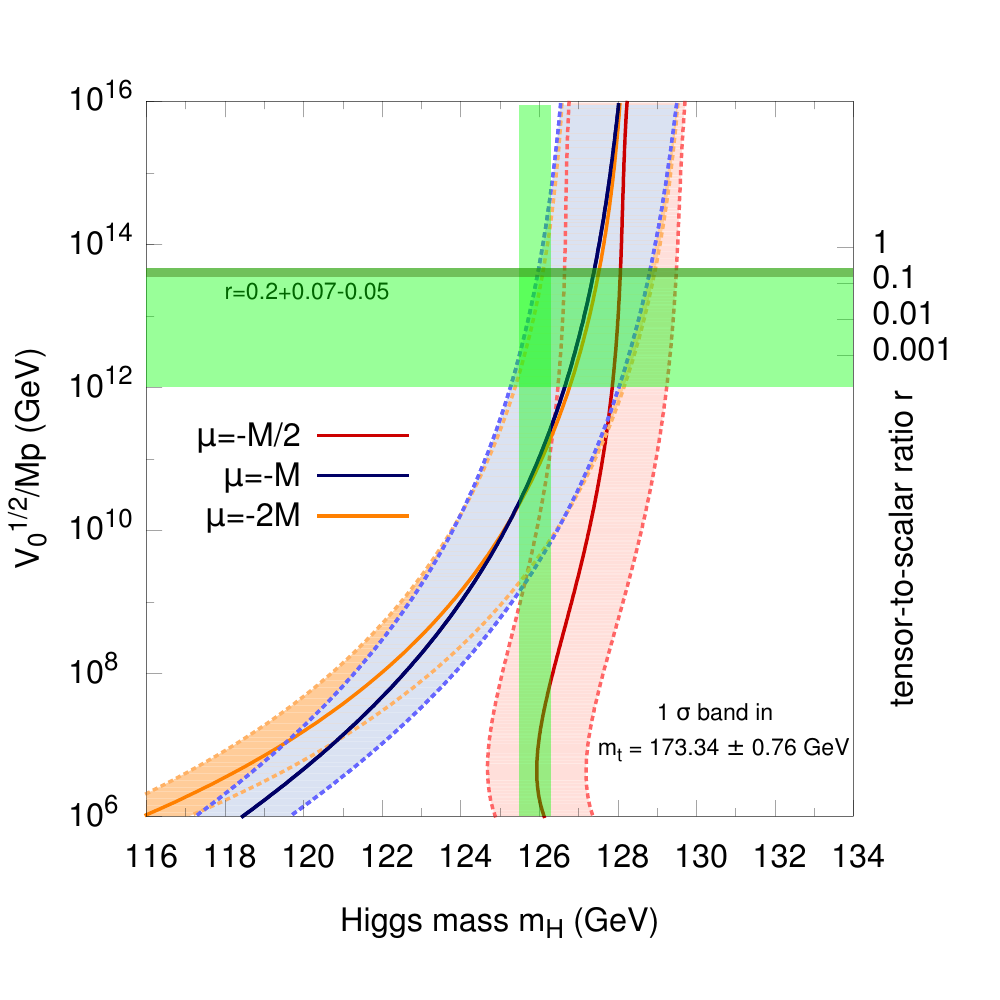}
\end{center}
\vspace{-0.5cm}
\caption{\label{gcu}  
The scale $M_{ss}=m_I\simeq V_0^{1/2}/M_p$ versus the Higgs mass  computed for three values of $\mu$. 
The bands correspond to results using the one sigma values for $m_{top}=173.3\pm 0.7$
from the LHC+Fermilab average \cite{topmass}.
The vertical band gives the measured Higgs mass at LHC \cite{higgsmass}. The thin horizontal line 
corresponds to the SUSY breaking scale $V_0^{1/2}/M_p$ obtained from the
observed tensor ratio $r$ using eq.(\ref{tomaya}). 
}
\label{plotguay}
\end{figure}

{\it Results.}
In the present case we are identifying $M_{ss}$ with the value of $V_0^{1/2}/M_p$ suggested by
BICEP2 data.   We have performed a computation of the value of the Higgs
mass under the assumption that this is the SUSY breaking scale and that $tan\beta =1$ at the unification scale.
The unification scale is fixed by identifying it with  the scale at which the  $g_2$ and $g_3$ SM gauge couplings unify,
assuming there are  threshold corrections which make them consistent also with unification with
$g_1$. 
The required threshold corrections  may come from a variety of sources. For example, if the $U(1)_Y$ slightly
mixes with a hidden $U(1)$ the hypercharge normalisation slightly changes in the correct direction, see
e.g. \cite{watari,RR}.
The resulting
unification scale is $M_X=10^{16}$, also consistent with the BICEP2 results for $V_0^{1/4}$. 
In our computation we have  performed 
the running of gauge and Higgs couplings at two loops, and the running of the soft terms in the
range $M_{ss}-M_X$  to one loop.  We also included  SM threshold corrections at the EW scale and
soft terms dependent threshold corrections at $M_{ss}$, see \cite{iv} for details of the required computation
as well as references.

The results obtained are shown in fig.1. The vertical band  shows  the $1\sigma$  values for $m_H=125.9\pm 0.4$ \cite{higgsmass}.
The horizontal thin band  corresponds the the SUSY breaking scale  $V_0^{1/2}/M_p$ computed from  eq.(\ref{tomaya})
in terms of the tensor ratio r.  The wider horizontal line extending it to $10^{12}$ GeV  allows from uncertainties from e.g.
dim$>5$ operators.  For the computation of  tan$\beta$ we have chosen for defniteness universal soft parameters 
$m_0^2=|M|^2/2, A=-3M/2$ with $M$ the gaugino mass, although the results are quite insensitive to the soft term structure. 
There is however a sizable dependence on the Higgs  $\mu$-parameter, and we display results for $\mu=-2M,-M, -M/2$
(the sign has negligible impact in the results though).
One sees that for a value of the top-quark within  $1\sigma$  of the LHC-Fermilab average \cite{topmass}, indeed 
{\it the Higgs mass is
consistent with the measured value of the potential energy at inflation being related to the SUSY breaking scale
as $M_{ss}=V_0^{1/2}/M_p$}.  This is the main result of this note.

{\it Final comments.}
If the scheme here proposed is correct, it would seem that the SUSY breaking scale would be about the
maximum  one compatible with the
restoration of  the Higgs potential stability. If $M_{ss}$ had been higher than $\simeq 10^{13}-10^{14}$ GeV, a Higgs minimum lower 
than the SM one would have developed.  But the fact that $M_{ss}$ is so high makes  the
inflation energy $V_0$ high enough  so as to leave a sizeable imprint in the tensor modes of the CMB.
We  would have been quite lucky, a  lower value for $M_{ss}$ would have made the tensor modes undetectable. 
So although such a large $M_{ss}$ scale  would have made SUSY undetectable at LHC, at least  it could have left
its imprint in the CMB.

Let us close this note with a few comments on additional implications of the existence of
a such large SUSY-breaking scale $M_{ss}\simeq 10^{12}-10^{13}$ GeV.
The fact that the extrapolated Higgs boson self-coupling $\lambda$ approaches zero (if one includes 
$2\sigma$ errors for the top-quark mass) not far from the Planck scale, and that the correspondding $\beta$ function is
also numerically  close to zero at those scales,  has been suggested as a hint for a conformal symmetry at the
Planck scale \cite{piraos}.  The observation of BICEP2 points to a new fundamental mass scale below $M_p$, 
making  that possibility unlikely. On the other hand
large intermediate scales have been considered in particle physics in a variety of contexts. 
In particular a Majorana right-handed neutrino mass of the same order $\simeq 10^{12}-10^{13}$ GeV,
would be consistent with appropriate sea-saw neutrino masses  for the left-handed neutrinos.
In a different vein, in a scheme with such a large SUSY mass, the neutralinos are not
available to become the dark matter in the universe. A natural candidate in this situation would be an
axion.  In fact the BICEP2 measurements strongly constraint also the  allowed axion decay constant $f_a$.
 In particular high scale CDM axions with $f_a> 10^{14}$ GeV would be ruled out.
Such axions would create large isocurvature fluctuations which are severely constrained by 
Planck data \cite{bicepaxions}.  
 Finally, the existence of a mass scale $V_0\simeq (10^{16})^4 \text{GeV}^4\simeq M_X^4$,
with $M_X$ the unification scale,  makes plausible the generation of proton decay operators
which could lead to detectable signatures 
at underground
experiments.

While this  connection between the Higgs mass and the 
inflation scale is very attractive,  it remain mysterious how
a simple polynomial scalar potential with ultra-Planck field values can make sense  in a
putative ultraviolet  completion of the theory.  In particular, in the context of string theory there are two 
new mass scales which are the compactification scale $M_c$ and the string scale $M_s$.
In the simplest situations those two scales are very close and of order the unification scale,
$M_X \simeq M_c\lesssim M_s\ll M_p$
(see e.g. \cite{BOOK} for a discussion of these).  Above a scale of order $10^{16}$ GeV a
4D field theory no longer makes sense and one cannot ignore, at least in principle, the KK and  string
excitations. Thus the apparent success of  such simple field theory scalar potentials is somewhat
surprising in the string context.  
 The BICEP results are giving us invaluable information which hopefully will shed light
on the UV completion of the SM.





\begin{acknowledgments}

We thank  P.G. C\'amara, V. Diaz and  F. Marchesano for useful discussions.
This work has been supported by the ERC Advanced Grant SPLE under contract ERC-2012-ADG-20120216-320421 and 
by  the grants FPA 2009-09017, FPA 2009-07908, and FPA 2010-20807-C02.
We also thank the  spanish MINECO {\it Centro de excelencia Severo Ochoa Program} under grant SEV-2012-0249. I.V. is supported through the FPU grant AP-2012-2690.

\end{acknowledgments}



\begin{thebibliography}{99}


\bibitem{bicep2}
  P.~A.~R.~Ade {\it et al.}  [BICEP2 Collaboration],
  ``BICEP2 I: Detection Of B-mode Polarization at Degree Angular Scales,''
  arXiv:1403.3985 [astro-ph.CO].

\bibitem{chaotic}
  A.~D.~Linde,
  ``Chaotic Inflation,''
  Phys.\ Lett.\ B {\bf 129} (1983) 177.


\bibitem{veronica}
  X.~Calmet and V.~Sanz,
  ``Excursion into Quantum Gravity via Inflation,''
  arXiv:1403.5100 [hep-ph].


\bibitem{kl}
  R.~Kallosh and A.~D.~Linde,
  ``Landscape, the scale of SUSY breaking, and inflation,''
  JHEP {\bf 0412} (2004) 004
  [hep-th/0411011].
  
  \bibitem{aggj}
   L.~Alvarez-Gaume, C.~Gomez and R.~Jimenez,
  ``Initial conditions for inflation and the energy scale of SUSY-breaking from the (nearly) gaussian sky,''
  Cosmology and Particle Physics beyond Standard Models : Ten Years of the SEENET-MTP Network (2014) 1
   [
  [arXiv:1307.0696 [hep-th]].


\bibitem{manybicep}
  K.~Nakayama and F.~Takahashi,
  ``Higgs Chaotic Inflation and the Primordial B-mode Polarization Discovered by BICEP2,''
  arXiv:1403.4132 [hep-ph];
   A.~Kehagias and A.~Riotto,
  ``Remarks about the Tensor Mode Detection by the BICEP2 Collaboration and the Super-Planckian Excursions of the Inflaton Field,''
  arXiv:1403.4811 [astro-ph.CO];
    T.~Kobayashi and O.~Seto,
  ``Polynomial inflation models after BICEP2,''
  arXiv:1403.5055 [astro-ph.CO];
    K.~Harigaya and T.~T.~Yanagida,
  ``Discovery of Large Scale Tensor Mode and Chaotic Inflation in Supergravity,''
  arXiv:1403.4729 [hep-ph];
  K.~Harigaya, M.~Ibe, K.~Schmitz and T.~T.~Yanagida,
  ``Dynamical Chaotic Inflation in the Light of BICEP2,''
  arXiv:1403.4536 [hep-ph];
  M.~P.~Hertzberg,
  ``Inflation, Symmetry, and B-Modes,''
  arXiv:1403.5253 [hep-th];
  S.~Choudhury and A.~Mazumdar,
  ``Reconstructing inflationary potential from BICEP2 and running of tensor modes,''
  arXiv:1403.5549 [hep-th].
  

\bibitem{bicepaxions}
T.~Higaki, K.~S.~Jeong and F.~Takahashi,
  ``Solving the Tension between High-Scale Inflation and Axion Isocurvature Perturbations,''
  arXiv:1403.4186 [hep-ph];
   L.~Visinelli and P.~Gondolo,
  ``Axion cold dark matter in view of BICEP2 results,''
  arXiv:1403.4594 [hep-ph];
    D.~J.~E.~Marsh, D.~Grin, R.~Hlozek and P.~G.~Ferreira,
  ``Tensor Detection Severely Constrains Axion Dark Matter,''
  arXiv:1403.4216 [astro-ph.CO].





\bibitem{hn}
  L.~J.~Hall and Y.~Nomura,
  ``A Finely-Predicted Higgs Boson Mass from A Finely-Tuned Weak Scale,''
  JHEP {\bf 1003} (2010) 076
  [arXiv:0910.2235 [hep-ph]].


\bibitem{previos}
  J.~A.~Casas, J.~R.~Espinosa and M.~Quir\'os,
  ``Improved Higgs mass stability bound in the standard model and implications
  for supersymmetry,''
  Phys.\ Lett.\  B {342} (1995) 171;
  [hep-ph/9409458];
  ``Standard Model stability bounds for new physics within LHC reach,''
  Phys.\ Lett.\  B {382} (1996) 374.
  [hep-ph/9603227];
  G.~Isidori, G.~Ridolfi and A.~Strumia,
  ``On the metastability of the standard model vacuum,''
  Nucl.\ Phys.\ B\ {609} (2001) 387
  [hep-ph/0104016]



\bibitem{elias}
J.~Elias-Miro, J.~R.~Espinosa, G.~F.~Giudice, G.~Isidori, A.~Riotto and A.~Strumia,
  ``Higgs mass implications on the stability of the electroweak vacuum,''
  Phys.\ Lett.\ B {\bf 709} (2012) 222
  [arXiv:1112.3022 [hep-ph]];
  G.~Degrassi, S.~Di Vita, J.~Elias-Miro, J.~R.~Espinosa, G.~F.~Giudice, G.~Isidori and A.~Strumia,
  ``Higgs mass and vacuum stability in the Standard Model at NNLO,''
  JHEP {\bf 1208} (2012) 098
  [arXiv:1205.6497 [hep-ph]]; D.~Buttazzo, G.~Degrassi, P.~P.~Giardino, G.~F.~Giudice, F.~Sala, A.~Salvio and A.~Strumia,
  ``Investigating the near-criticality of the Higgs boson,''
  JHEP {\bf 1312} (2013) 089
  [arXiv:1307.3536];  
  
 
 \bibitem{hebecker1}
  A.~Hebecker, A.~K.~Knochel and T.~Weigand,
  ``A Shift Symmetry in the Higgs Sector: Experimental Hints and Stringy Realizations,''
  JHEP {\bf 1206}, 093 (2012)
  [arXiv:1204.2551 [hep-th]].

\bibitem{imrv}
   L.~E.~Ib\'a\~nez, F.~Marchesano, D.~Regalado and I.~Valenzuela,
  ``The Intermediate Scale MSSM, the Higgs Mass and F-theory Unification,''
  JHEP {\bf 1207} (2012) 195
  [arXiv:1206.2655 [hep-ph]].



\bibitem{iv}
  L.~E.~Ib\'a\~nez and I.~Valenzuela,
  ``The Higgs Mass as a Signature of Heavy SUSY,''
  JHEP {\bf 1305} (2013) 064
  [arXiv:1301.5167 [hep-ph]].


\bibitem{hebecker2}
  A.~Hebecker, A.~K.~Knochel and T.~Weigand,
  ``The Higgs mass from a String-Theoretic Perspective,''
  Nucl.\ Phys.\ B {\bf 874} (2013) 1
  [arXiv:1304.2767 [hep-th]].








\bibitem{otrosinter}
C. Liu, Z.-h. Zhao,
  ``$\theta_{13}$ and the Higgs mass from high scale supersymmetry,''
  Commun.\ Theor.\ Phys.\  {\bf 59} (2013) 467
  [arXiv:1205.3849 [hep-ph]].;
   L.~J.~Hall and Y.~Nomura,
  ``Grand Unification and Intermediate Scale Supersymmetry,''
  JHEP {\bf 1402} (2014) 129
  [arXiv:1312.6695 [hep-ph]];
    M.~Ibe, S.~Matsumoto and T.~T.~Yanagida,
  ``Flat Higgs Potential from Planck Scale Supersymmetry Breaking,''
  arXiv:1312.7108 [hep-ph].
 


 \bibitem{split}
  N.~Arkani-Hamed and S.~Dimopoulos,
  ``Supersymmetric unification without low energy supersymmetry and signatures for fine-tuning at the LHC,''
  JHEP {\bf 0506} (2005) 073
  [hep-th/0405159].
  
  
  

\bibitem{topmass}
``First combination of Tevatron and LHC measurements of the top-quark mass'',
ATLAS, CDF, CMS and D0 collaborations,
arXiv:1403.4427[hep-ex].




\bibitem{higgsmass}
Particle Data Group, 
http://pdg.lbl.gov/2012/reviews/r\\
pp2012-rev-higgs-boson.pdf.


  

\bibitem{watari}
  R.~Tatar and T.~Watari,
  ``GUT Relations from String Theory Compactifications,''
  Nucl.\ Phys.\ B {\bf 810} (2009) 316
  [arXiv:0806.0634 [hep-th]].


\bibitem{RR}
  P.~G.~C\'amara, L.~E.~Ib\'a\~nez and F.~Marchesano,
  ``RR photons,''
  JHEP {\bf 1109} (2011) 110
  [arXiv:1106.0060 [hep-th]].


\bibitem{piraos}
  F.~Bezrukov, M.~Y.~Kalmykov, B.~A.~Kniehl and M.~Shaposhnikov,
 ``Higgs boson mass and new physics,''
  [hep-ph/1205.2893];
D.L.~Bennett, H.B.~Nielsen and I.~Picek, Phys.Lett.B {208}{1988}{275};
C.D.~Froggatt and H.B.~Nielsen, Phys.Lett.B {368}{1996}{96};
  M.~Shaposhnikov and C.~Wetterich,
  ``Asymptotic safety of gravity and the Higgs boson mass,''
  Phys.\ Lett.\ B {683} (2010) 196
  [hep-ph/0912.0208];
  M.~Holthausen, K.~S.~Lim and M.~Lindner,
  ``Planck scale Boundary Conditions and the Higgs Mass,''
  JHEP {1202} (2012) 037
  [arXiv:1112.2415].



\bibitem{BOOK}
L.E. Ib\'a\~nez  and  A. Uranga,
{\it String Theory and Particle Physics. An Introduction
to String Phenomenology}. Cambridge U.P., (2012).



\end{thebibliography}
\end{document}